\documentclass[aps,twocolumn,superscriptaddress,floatfix,amsmath,a4paper]{revtex4}
\usepackage[utf8]{inputenc}
\usepackage{bm,amsmath,amssymb,mathtools,color,graphicx}
\usepackage{textcomp,gensymb,units}
\newcommand{\D}{\mathrm{d}}
\newcommand{\E}{\mathrm{e}}

\newcommand{\NC}{N_{\scriptscriptstyle\rm C}}
\newcommand{\ND}{N_{\scriptscriptstyle\rm D}}

\begin{document}
%%%%%%%%%%%%%%%%%%%%%%%%%%%%%%%%%%%%%%%%%

\title{Emergent cooperative behavior in transient compartments}

\author{Jeferson J. Arenzon}
\email{arenzon@if.ufrgs.br}
\affiliation{Instituto de F\'\i sica, Universidade Federal do
Rio Grande do Sul, CP 15051, 91501-970, Porto Alegre RS, Brazil}
\affiliation{Instituto Nacional de Ciência e Tecnologia - Sistemas
Complexos, Rio de Janeiro RJ, Brazil}

\author{Luca Peliti}
\email{luca@peliti.org}
\affiliation{Santa Marinella Research Institute, 00058, Santa Marinella (RM), Italy}
\date{\today} 

\begin{abstract}
We introduce a minimal model of multilevel selection on structured populations, considering the interplay between game theory and population dynamics. 
Through a bottleneck process, finite groups are formed with cooperators and defectors sampled from an infinite pool.
After the fragmentation, these  transient compartments grow until the maximal number of
individuals per compartment is attained.
Eventually, all compartments are merged,  well mixed and the whole process is repeated.
We show that cooperators, even if interacting only through mean-field intra-group interactions that favor defectors, may perform well because of the inter-group competition and the size diversity among the compartments.
These cycles of isolation and coalescence may therefore be  important in maintaining diversity among different species or strategies and may help to understand the underlying mechanisms of the scaffolding processes in the transition to multicellularity.
\end{abstract}

\maketitle

\section{Introduction}

Cooperation may emerge from the selfish behavior of individuals. 
This long-standing evolutionary paradox~\cite{Axelrod84} has been thoroughly studied, unveiling several possible mechanisms that seed and sustain cooperative behavior~\cite{Nowak06b,SzFa07,Frey10,Percetal2013,AdScHi16}. 
Network reciprocity, among them, depends on the spatial correlations generated from continued interactions~\cite{Axelrod84,NoMa92}, but not necessarily on complex strategies and behaviors, or high-level cognitive skills.
Indeed, cooperators (C) may thrive because of the presence of persistent groups, where the mutual protection from other cooperators in the bulk outweighs the exploitation by defectors (D) on the surface.
In the absence of such spatial structure, particularly when the population is well-mixed, defectors  have a larger average fitness~\cite{Smith82,HoSi98} and other mechanisms must be in place in order to maintain cooperation~\cite{Nowak06b}.

Natural selection may operate across different scales of organization, from the individual-level to higher orders that involve groups of agents, sometimes isolated in different, temporary compartments.
These within-group and between-groups interactions may be relevant for studying structured populations~\cite{HaHoDo06,Tarnita17,Matsumura16,BlLaNgPe18,BlNgPeLa20,CoLeMoPl23} and community coalescence~\cite{Gilpin94,Rillig15,Tikhonov16,AnVeRi19,Dutton21,LeClCoSmPa21,Colunga22}, in particular the interplay between competition and cooperation.
There are also similarities with patchy systems where environmental factors have a major influence, e.g., temporary pools~\cite{Souza79,BlSc01,Veach16,XuDy17}, and bottleneck processes, alternating between low-number populations and a growth phase, that are common in many microbial cycles and other processes~\cite{NaDrFo16,NiBlBoRa23}, being amenable to be experimentally probed~\cite{GrWeBu04,ChRiLe09}.
Similar ingredients have been considered in the early modelling of group and multilevel selection (see, e.g., \cite{Wynne63,Wilson75,Smith76,WiWi08} and references therein).
Also, social individuals interact during their lifetimes with several groups, in a succession of different social environments leading to collective adaptation.
Another instance of changing environments occurs when agents are mobile~\cite{VaSiAr07,MeBuFoFrGoLaMo09,VaBrAr14,Javarone16}.
Different models have considered transient compartments (or groups) and studied its benefit to cooperating agents, sustaining diversity of strategies and the coexistence with defectors.

We study here a minimal model for multilevel selection, where isolated compartments are created and populated with individuals taken from a large (infinite) pool.
The random initial size and composition of each of these transient compartments determine how their population grow and how the overall fraction of cooperators gets modified.
The interior of each compartment has no structure, agents are fully-mixed and the dynamics is mean-field, given by the replicator equation.
%Inside each of these compartments, the agents interact, reproduce and, eventually are merged once again in a common pool before the whole process is repeated. The population is structured albeit the compartments are transient since 
After attaining their maximal number of individuals, all compartments coalesce and their populations becomes, once again, fully mixed.
This whole sequence of fragmentation, growth and merging is repeated and may lead to sustained cooperative behavior~\cite{CrMeFr12,Steiner21}.
These fluctuations in the number of individuals are important in the fixation and extinction mechanisms of strategies or traits, bringing together game theory and population dynamics~\cite{HaHoDo06,HuHaTr15}.
Compartments where both cooperators and defectors are initially included, because of the unstructured internal interactions, will be dominated by the latter.  
The main idea, instead, is that compartments in which cooperators are prevalent will have a larger population at the end of the growth phase, and therefore will contribute more to the total pool population.
Thus, depending on the conditions, cooperators may thrive: the underlying mechanism is akin to the Simpson's Paradox~\cite{Simpson51,Blyth72,HaMoHoSi02,HaHoDo06,ChRiLe09,CrMeFr12,GeBr20}.

%Indeed, the result is equivalent to an average over many different outcomes of a steep decrease of a population and its consequent loss of diversity.

The fundamental motivation to study this class of models is to better understand the role of the compartmental structure and its different levels of selection in scaffolding stable multicellular groups. 
Despite multicellularity being advantageous once established, the first steps leading to a higher-level organisms correspond to a major evolutionary transition~\cite{SmSz97},  shifting the level at which natural selection operates and probably involve transient coordination between cooperating individuals~\cite{Jacobeen18}.
Our minimal model addresses the above general questions and also, in particular, which is the role of  transient compartments in the transition to multicellularity.
It shows that cooperation is possible with mean-field internal interactions and without permanent groups of agents, where defection is known to usually prevail.
In the next sections we describe our simple framework with temporary clustering and study the long-term behavior of the population of cooperators.

\section{Transient Compartments}

\subsubsection{The Prisoner's Dilemma Game}

Cooperation involves a cost $c>0$ for the provider agent (C) while another individual receives the benefit $b$. 
Instead, defection (D) has no cost whatsoever. 
Thus, within the usual parametrization of the Prisoner's Dilemma game (PDG), two interacting cooperators get a reward $b-c$ each ($b>c$), while two defectors are punished and receive nothing. 
When a D meets a C, the former payoff is $b$, while the cost is $-c$ for the latter. 
The interaction between two defectors does not involve any benefits or costs.
%For the Prisoner's Dilemma, $c>0$, while for the Snowdrift, $c<0$. 
Summarizing:
\begin{center}
\begin{tabular}{c|ccc}
&C& &D\\
\hline
C&$b-c$& & $-c$\\
D&$b$& & 0
\end{tabular}
\end{center}

\begin{figure}[htb]
  \includegraphics[width=0.9\columnwidth]{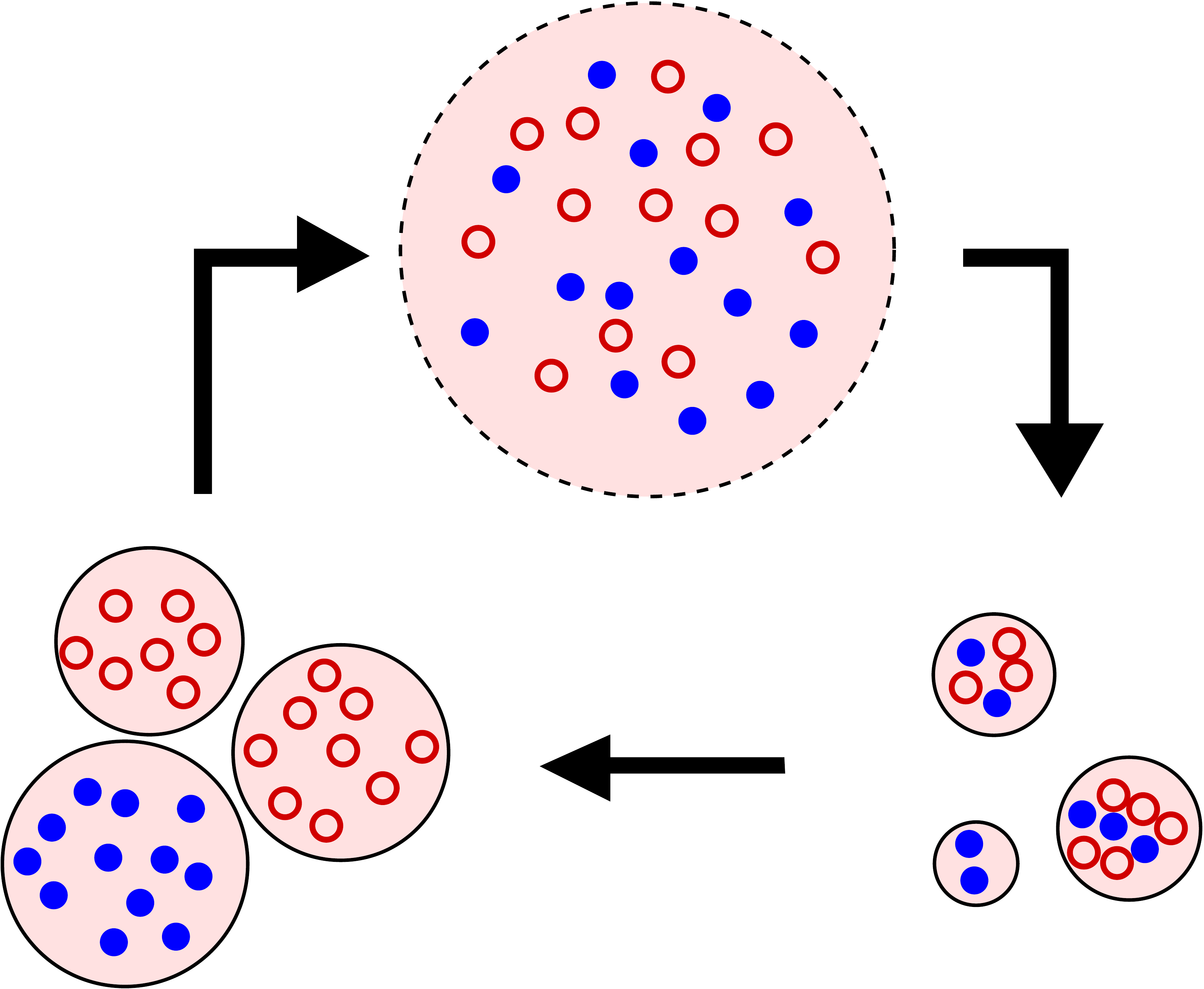}
\caption{Schematic representation of the model. Cooperators (blue/filled) and defectors (red/open) are initially fully mixed in an infinite pool with concentrations $x$ and $1-x$, respectively (top). Finite compartments are created and randomly populated (bottom right). Each one evolves independently (growth phase, see text) and attains a stationary state (bottom left). Cooperators disappear from any compartment with a mixed initial state, but their initial number determines the final number of defectors. Only all-C compartments can sustain cooperation.  All compartments are eventually merged back in a common pool (coalescence phase), densities are updated and the cycle is repeated.}
\label{fig.ciclo}
\end{figure}

We consider a infinitely large pool containing a mixture of cooperators (C) and defectors (D), with initial proportions $x$ and $1-x$, respectively (see Fig.~\ref{fig.ciclo}). 
From this pool, random samples are used to seed infinitely many isolated compartments. 
In the {\it growth phase}, each compartment evolves independently 
%following the usual mean-field PDG dynamics, 
until resources (i.e., the accumulated benefits from cooperators) are exhausted and the maximal number of individuals is attained. 
The population from all compartments are then merged ({\it coalescence phase}), forming a new pool with an updated fraction $x'$ of cooperators, and the cycle starts again until a fixed point, $x=x'\equiv x^*$, is attained. 
Our aim is to verify whether cooperators survive in the long term ($x^*>0$), either dominating the whole system ($x^*\to 1$) or somehow coexisting with defectors ($0<x^*<1$), along with the respective intervals of the parameters in which the different phases occur. 

\subsubsection{Agent-level dynamics: growth phase}

Assuming that agents inside a compartment are fully mixed, the average payoff of a cooperator is the benefit received from the other cooperators, minus the cost of benefiting all agents, i.e., $f_{\scriptscriptstyle\rm C} = b\rho-c$, where $\rho$ is the fraction of cooperators in the given compartment, that may differ, due to its finite size, from the the density in the infinite pool. Defectors receive, instead, $b$ from each cooperator and $f_{\scriptscriptstyle\rm D} = b\rho$. 
%\begin{align*}
%  f_{\scriptscriptstyle\rm C} &= bx-c \\ %(b-c)x -c(1-x) \\
%  f_{\scriptscriptstyle\rm D} &= bx.
%\end{align*}
The average fitness (that we consider to scale with the payoff) is $F(\rho)=(b-c)\rho$ and
%\begin{equation}
%F(x) = (b-c)x^2-c(1-x)x+bx(1-x) = (b-c)x
%\end{equation}
the mean-field, replicator dynamics is given by~\cite{HoSi98}
\begin{align}
  \dot{\rho} = \rho(f_{\scriptscriptstyle\rm C}-F)=-c\rho(1-\rho).
\label{eq.replicator}
\end{align}
We consider demographic variations allowing the total number $N$ of agents in each compartment to fluctuate. 
In each compartment, the population depends on the instantaneous average fitness through $\dot{N}=F(\rho)N$.
Thus, the asymptotic population is $N_{\infty}= N_0 \exp{[\Phi(\rho_0)]}$, where $\Phi(\rho_0)$ is the accumulated fitness
\begin{align*}
%\begin{split}
  \Phi(\rho_0)&=\int_0^{\infty}\D t\;F(\rho)=-\int_{\rho_0}^{\rho_{\infty}}\D \rho\;\frac{F(\rho)}{c \rho (1-\rho)}\nonumber\\
  %&=-\int_{x_{0}}^{x_{\infty}}\D x\;\frac{b-c}{c(1-x)}
  &=\frac{b-c}{c}\,  \ln\frac{1-\rho_{\infty}}{1-\rho_0}, %\Delta \ln (1-x),
%\end{split}
\end{align*}
and $\rho_0$ and $\rho_{\infty}$ are, respectively, the initial and the asymptotic fraction of cooperators in the compartment.
Notice that $\Phi(\rho_0)$ has a logarithmic divergence as $\rho_{0}\to 1$ as the few remaining defectors benefit much from the large number of cooperators. 
The eventual size of a compartment with $\rho_0<1$ (for compartments with only cooperators, $\rho_0=1$, see below) where all remaining individuals are defectors ($\rho_{\infty}=0$), is
\begin{equation}
N_{\infty}= %N_{0}\,\E^{\phi(x_{0})}=
\frac{N_0}{\left(1-\rho_0\right)^{b/c-1}}.
\label{eq.Ninf}
\end{equation}
%We assume that a compartment with only cooperators, $\rho_0=1$, grows to a maximum size $N_{\max}$, independently of the initial size (see below). 

The seed of a compartment initially contains $n$ agents, of which $m$ are cooperators, drawn with probability $x$ from the infinite pool. 
Hence $\rho_0=m/n$, and unless all agents are cooperators (i.e., as long as $\rho_0<1$) the compartment eventually contains, from Eq.~(\ref{eq.Ninf}), $N_{n,m}=n(1-m/n )^{1-b/c}$ agents, all defectors.
The average number of defectors per compartment after the growth phase is given by
\begin{align}
N_{\scriptscriptstyle\rm D}(x)&=\sum_{n=1}^{\infty} \sum_{m=0}^{n-1}p_{n,m}N_{n,m}
\nonumber
\\
&=\sum_n \frac{1}{n!}\frac{ \lambda^n n^{b/c}}{\E^{\lambda}-1} 
   \sum _{m<n}\!\binom{n}{m} \frac{(1-x)^{n-m} x^m}{(n-m)^{b/c-1}} ,
\label{eq.ND}   
\end{align}
where $p_{n,m}$ %=\text{Poisson}(\lambda, n)\,\text{Binomial}(1-x, n, m)$ 
is the product of a zero-truncated Poisson distribution with parameter $\lambda$ and a binomial with parameters $n$ and $x$.
The $m=n$ term is not included because defector-free compartments do not contribute to $N_{\scriptscriptstyle\rm D}$.
The parameter $\lambda$ sets the average size of the compartments at their creation. %, $\lambda/(1-\E^{-\lambda})$. 
In the limit $x\to 0$, where no cooperator is initially present and there is no growth,   $N_{\scriptscriptstyle\rm D}(0)$ is equal to the average initial size of the compartments,  $N_{\scriptscriptstyle\rm D}(0)=\lambda/(1-\E^{-\lambda})$.

We consider that a compartment whose initial composition has only cooperators increases to size $N_{\max}$, independently of the initial size $n$. 
After the growing phase, the average number of cooperators per compartment is 
\begin{align}
  N_{\scriptscriptstyle\rm C}(x)=\sum_{n=1}^{\infty} p_{n,n}N_{\max}%=N_{\max}\sum_{n=1}^{\infty}\frac{\E^{-\lambda} \lambda^{n}}{n!}x^{n}\nonumber\\
  =N_{\max}\frac{\E^{\lambda x}-1}{\E^{\lambda}-1}.
  \label{eq.NC}
\end{align}
In the limit $x\to 1$, all clusters become purely cooperative and grow to the maximum size, $N_{\scriptscriptstyle\rm C}=N_{\max}$.
It is important to emphasize that cooperator survival is only due to these defector-free compartments. Since no compartment size can exceed $N_{\max}$, the size of compartments with defectors ($\rho<1$) at the end of the growth phase is given by $\min\left(N_{\mathrm{D}}(x),N_{\max}\right)$, where $N_{\mathrm{D}}(x)$ is given by Eq.~(\ref{eq.ND}).
We also assume that $\lambda\ll N_{\max}$.

\subsubsection{Group-level dynamics: coalescence phase}

Finally, in the last step of the dynamics, all compartments are merged into a single, fully mixed pool and the result of a full cycle is summarized by the mapping $x\rightarrow x'$, where
\begin{equation}
x'=\frac{N_{\scriptscriptstyle\rm C}(x)}{N_{\scriptscriptstyle\rm C}(x)+N_{\scriptscriptstyle\rm D}(x)}.
\label{eq.map}
\end{equation}
This sequence is repeated until a fixed point $x=x'\equiv x^*$ is attained.
Along with the stability condition $dx'/dx|_{x^*}\leq 1$ it determines the asymptotic behavior of the system. 
In the next section we explore the possible outcomes for different values of the parameters.

\section{Results}

\subsection{Fixed points and stability}
\label{sec.original}

\begin{figure}[htb]\includegraphics[width=\columnwidth]{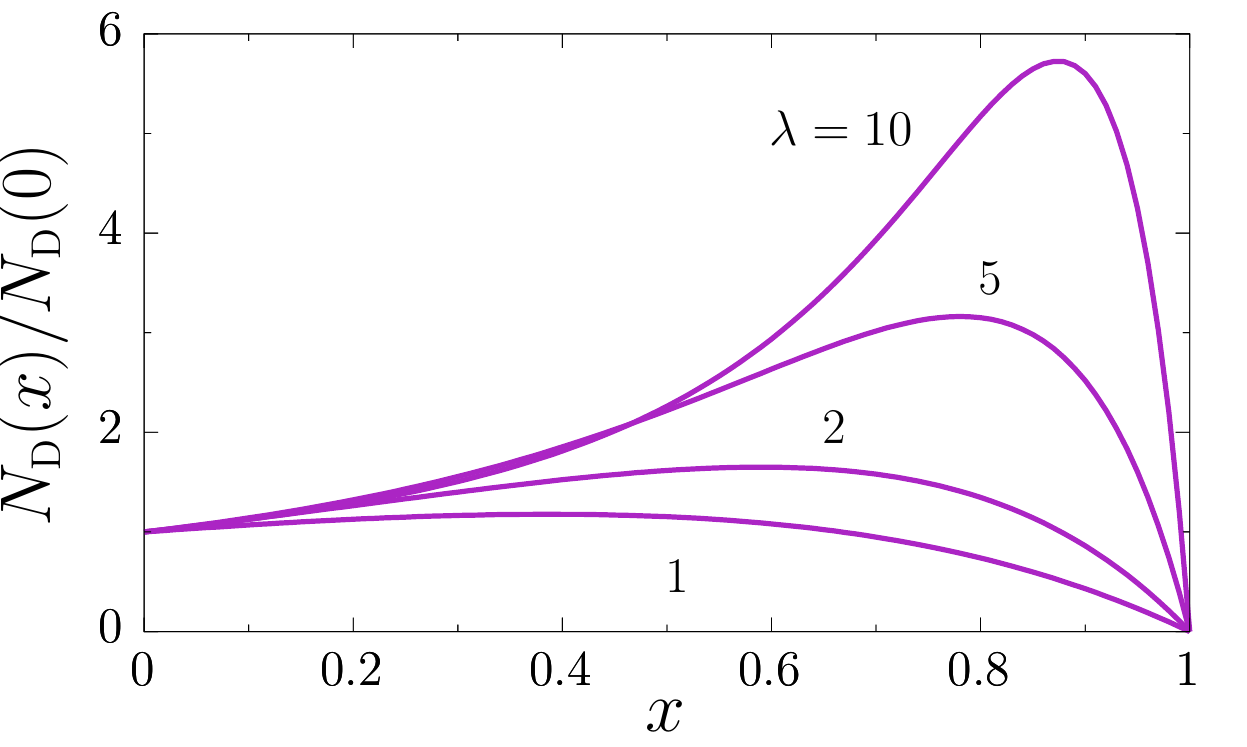}
\caption{Average number of defectors per compartment, $N_{\scriptscriptstyle\rm D}(x)$, Eq.~(\ref{eq.ND}), rescaled by $N_{\scriptscriptstyle\rm D}(0)=\lambda/(1-\E^{-\lambda})$,  as a function of the fraction $x$ of cooperators for $b/c=2$. Defectors increase in number until a $\lambda$-dependent maximum is attained and then disappear as $x\to 1$.}
\label{fig.ND}
\end{figure}

The asymptotic average number of defectors per compartment, Eq.~(\ref{eq.ND}), is shown in Fig.~\ref{fig.ND} as a function of $x$ for several values of the parameter $\lambda$. 
As mentioned before, in the limit $x\to 0$ there is no growth and $N_{\scriptscriptstyle\rm D}$ is the average initial size of the compartments,  $N_{\scriptscriptstyle\rm D}(0)=\lambda/(1-\E^{-\lambda})$.
As $x$ increases, defectors exploit the cooperators in the same compartment and, consequently, their total number increases. 
On the other hand, for larger values of $x$, those pools that are fully occupied by cooperators will hinder the overall growth of defectors. 
Because of these competing mechanisms (the exploitation of cooperators that increases the number of defectors and defector-free pools that favors cooperators), $N_{\scriptscriptstyle\rm D}$ presents a maximum at intermediate values of $x$.
When the average compartment size increases, following $\lambda$, the peak of $N_{\scriptscriptstyle\rm D}$ becomes higher and moves to larger values of $x$.
In other words, to obtain defector-free compartments one needs larger values of~$x$ when $\lambda$ is larger.

\begin{figure}[htb]
\begin{center}
  \includegraphics[width=\columnwidth]{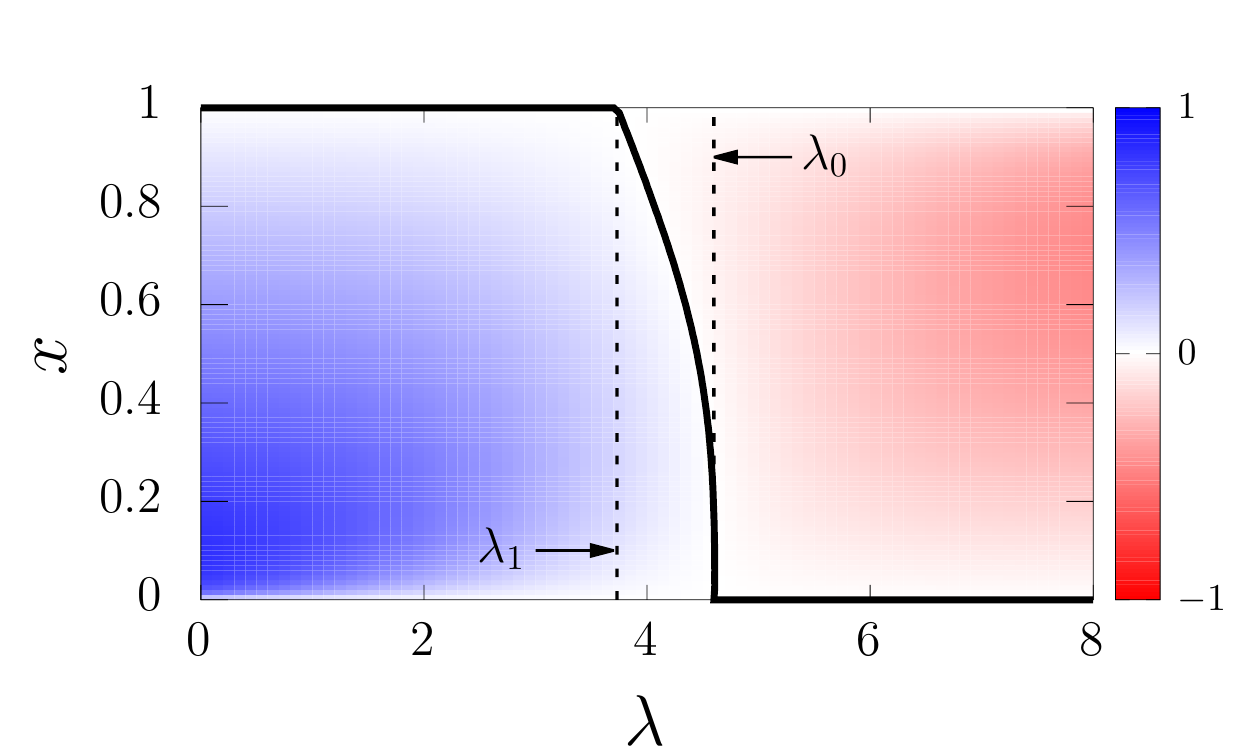}
\end{center}
\caption{The difference $\Delta x\equiv x'-x$ from Eq.~(\ref{eq.map}) for each point in the plane $(\lambda,x)$ is represented by the color code for the particular case $b/c=2$ and $N_{\max}=100$. The fixed points $x^*\equiv x=x'$ are indicated by the black isoclines. If the fraction of cooperators increases ($\Delta x>0$), the color is blue, while red indicates $\Delta x<0$ and an increase of defectors. In the interval $\lambda_1\leq\lambda\leq\lambda_0$, between the vertical dashed lines, only the coexistence solution $0<x^*<1$ is stable (see the phase diagram in Fig.~\ref{fig.phasediagram} for other values of $b/c$ and the possibility of a bistable phase as well).}
\label{fig.alpha1z}
\end{figure}

\begin{figure}[htb]
  \includegraphics[width=\columnwidth]{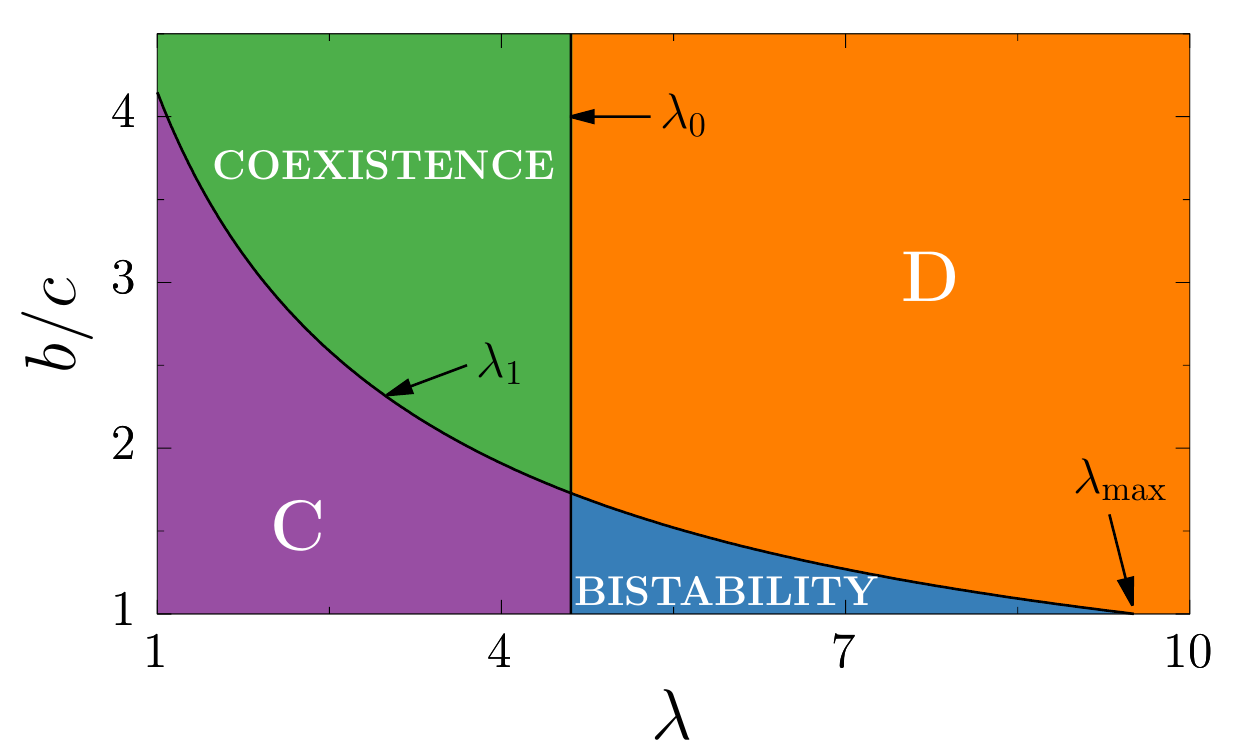}
\caption{Phase diagram in the plane $(\lambda, b/c)$ for $N_{\max}=100$.
%and unconstrained compartment sizes ($n_{\max}\to\infty$). 
The pure phases C and D have, respectively, $x^*=1$ and 0. These two solutions also occur, depending on the initial condition, in the bistable phase. 
Finally, in the coexistence phase, both cooperators and defectors are eventually present, $0<x^*<1$. 
The borders between the phases are obtained from the stability conditions: the vertical line is $\lambda=\lambda_0$ while $\lambda_1$ is the solution of Eq.~(\ref{eq.stability}) for each value of $b/c$. Beyond $\lambda_{\max}\simeq 9.51$ there is only the D phase. }
%The transition is continuous from the C to the coexistence phase while from the bistable to the D phase it is discontinuous.}
\label{fig.phasediagram}
\end{figure}

The behavior of $\Delta x = x'-x$ as a function of~$x$ and~$\lambda$ is shown in Fig.~\ref{fig.alpha1z}. Regions where $\Delta x >0$ are marked in blue, and those in which $\Delta x<0$ are marked in red. The line of fixed points $x^*(\lambda)$ is plotted as a solid black line
%Fig.~\ref{fig.alpha1z} shows an example of the behavior of $\Delta x=x'-x$, from Eq.~(\ref{eq.map}), in the plane $(\lambda,x)$ for $b/c=2$ and $N_{\max}=100$. 
%Positive, increasing values  are shown in blue while negative ones are in red. 
%The fixed points $x^*$, representing some of the possible phases, are the isoclines $x'=x$ and are indicated by a solid black line in the middle of the white interfaces. 
For small values of $\lambda$, the fixed point is $x^*=1$ and only cooperators survive (C phase). 
Before entering the large-$\lambda$ region where defectors fully dominate ($x^*=0$, D phase), there is an intermediate region for $\lambda_1<\lambda<\lambda_0$ (see below) where both strategies coexist, $0<x^*<1$. 
In addition to the three behaviors shown in this example, there is also a bistable phase in which, depending on the initial condition, the asymptotic state may be either $x^*=0$ or 1.
The four possible  phases, and the regions in which each one is stable in the plane $(\lambda,b/c)$, are shown in Fig.~\ref{fig.phasediagram} and discussed below.

The solution with cooperators fully populating the system, $N_{\scriptscriptstyle\rm D}=0$ and $x^*=1$, albeit always existing, is stable only for $\lambda\leq\lambda_1$ where $\lambda_1$ is obtained, for each value of $b/c$, from the stability threshold
\begin{align}
	 \left.\frac{dx'}{dx}\right|_{x^*=1} &= \frac{1}{N_{\max}} \frac{1}{\E^{\lambda_1}-1}\sum_{n=1}^{\infty} 
\frac{\lambda_1^n n^{b/c}}{(n-1)!} = 1.
\label{eq.stability}
\end{align}
For the example shown in Fig.~\ref{fig.alpha1z}, the solution $x^*=1$ is stable for $\lambda\leq\lambda_1\simeq 3.7$ (left vertical dashed line) where the coexistence region, with $0<x^*<1$, starts.

The phase diagram shown in Fig.~\ref{fig.phasediagram} exhibits four phases: C (only cooperators survive, $x^*=1$), D (only defectors survive, $x^*=0$), coexistence ($0<x^*<1$), and bistability ($x^*=0$ or 1 depending on the initial condition).
From the condition $dx'/dx|_{x^*=0}\leq 1$, we obtain that the solution $x^*=0$ is stable in the region $\lambda\geq\lambda_0\equiv\ln N_{\max}$, for all values of $b/c$.  
In the case of Fig.~\ref{fig.alpha1z}, $\lambda_0\simeq 4.6$ and is indicated  by the vertical dashed line on the right.
The lines $\lambda_0$ and $\lambda_1$ separate the different regions depicted in the phase diagram Fig.~\ref{fig.phasediagram} mentioned above.
The C phase
%, a small initial amount of cooperators always invade and populate the whole system (fixation), $N_{\scriptscriptstyle\rm D}=0$ and $x^*=1$.
is stable for $1\leq b/c\leq\lambda_1$ and $1\leq\lambda\leq\lambda_0$, the low, left region of Fig.~\ref{fig.phasediagram}.
This stability region, bounded by the value of $\lambda_1$, becomes smaller as $\lambda$ approaches $\lambda=\lambda_0$.
When $1< b/c < \lambda_1$ and $\lambda_0 < \lambda < \lambda_1$, both $x^*=0$ and $x^*=1$ are stable. Thus the asymptotic behavior is dictated by the initial conditions.
%For $\lambda\geq\lambda_0$, although $x^*=1$ remains stable, it shares the space of initial states with the $x^*=0$ solution while $1\leq b/c\leq\lambda_1$.
%In this bistable phase, both solutions coexist and the asymptotic state, whether $x^*=0$ or 1, depends on the initial configuration.
The phase D, in which $x^*=0$, appears for $\lambda>\lambda_0$ and $b/c>\lambda_1$. 
Both pure solutions become unstable for $\lambda<\lambda_0$ and $b/c>\lambda_1$.
In this region, coexistence between cooperators and defectors is possible and the solution $0<x^*<1$ becomes stable. 

The only region of the phase diagram Fig.~\ref{fig.phasediagram} in which cooperators are fully absent, the D phase, occurs for large compartments ($\lambda>\lambda_0$) and benefit-to-cost ratios ($b/c>\lambda_1$).
When the average compartment size is small enough, small groups dominated by cooperators allow their long-term survival. 
In the bistable region, where the compartments are larger, cooperators may invade all compartments for  initial states with a large fraction of cooperators.
Since $b/c$ cannot be smaller than~1, there is a maximum value, $\lambda_{\max}$, of $\lambda$ beyond which only the D phase persists. From Eq.~(\ref{eq.stability}), we obtain the following condition for $\lambda_{\max}$:
%From Eq.~(\ref{eq.stability}), the stability line  approaches the point $b/c=1$ for the value of $\lambda_1$ that satisfies the equation 
\begin{equation}
\lambda_{\max}(\lambda_{\max}+1)=N_{\max}[1-\exp(-\lambda_{\max})].
\label{eq.lmax}
\end{equation}
For $N_{\max}=100$, Fig.~\ref{fig.phasediagram}, we have $\lambda_{\max}\simeq 9.51$.
Thus, there is an average initial size for the compartments beyond which only the D phase exists, and this value behaves as $\lambda_{\max}\sim \sqrt{N_{\max}}$.

%The phase diagram Fig.~\ref{fig.phasediagram} presents an apparent paradox.
When the benefit of cooperating becomes large enough compared to its cost, the purely cooperating state becomes unstable.
For $\lambda<\lambda_0$, defectors are no longer completely removed (as in phase C) and may coexist with cooperators.
On the other hand, for $\lambda>\lambda_0$, defectors invade the whole system (phase D).
Thus, increasing $b/c$ is detrimental to cooperating agents, helping, instead, those that defect (equivalently, for a fixed $b$, increasing the cost $c$ for cooperators, enhances the possibility of cooperation).
Interestingly, even if whenever both defectors and cooperators are present in the same compartment, the former dominates, eliminating the cooperators, the overall population of cooperators may increase.
Those defector-free compartments indeed attain an even larger size as an effect of the mutual support. 
On the other hand, cooperators have a larger probability of survival as the average compartment size decreases.
These are the consequences of the non-trivial interplay between the rules of the PDG and the independent evolution of the isolated compartments, and the basic mechanism is akin to the Simpson's paradox~\cite{Simpson51,Blyth72,ChRiLe09,CrMeFr12,GeBr20}. 
%The benefits of cooperation are harvested both by cooperators and defectors.

%For example, for the particular value $b/c=2$, the above expression simplifies,
%$\lambda(1+3\lambda+\lambda^2)=N_{\max} (1-\E^{-\lambda})$,
%whose solution is $\lambda \simeq 3.73$ for $N_{\max}=100$ (otherwise, it goes as $\lambda\sim N_{\max}^{1/3}$ as $N_{\max}$ increases).

The above results considered that the initial compartment size is not constrained whatsoever. 
The poissonian probability used both for $\NC(x)$ and $\ND(x)$ (and in the corresponding stability relations) considers that all initial sizes $n$ for the compartments, from $n=1$ to $n\to\infty$, are taken into account.
However, some environments may constrain the compartments to have initial sizes smaller than a maximum value, $n\leq n_{\max}$.
This imposes an upper limit in the $n$-sum but does not, necessarily, constrain the growth phase and the value of $N_{\max}$. 
For $\lambda\ll n_{\max}$, such a cutoff has no relevant role and the results are equivalent to $n_{\max}\to\infty$.
In the other extreme, $\lambda\gg n_{\max}$, most of the compartments will have the maximum size and only the term $n=n_{\max}$ contributes to the sums.
This limiting case, where compartments have uniform sizes, is briefly discussed in the next section.

\section{Uniform-size Compartments}

\begin{figure}[htb]
\includegraphics[width=\columnwidth]{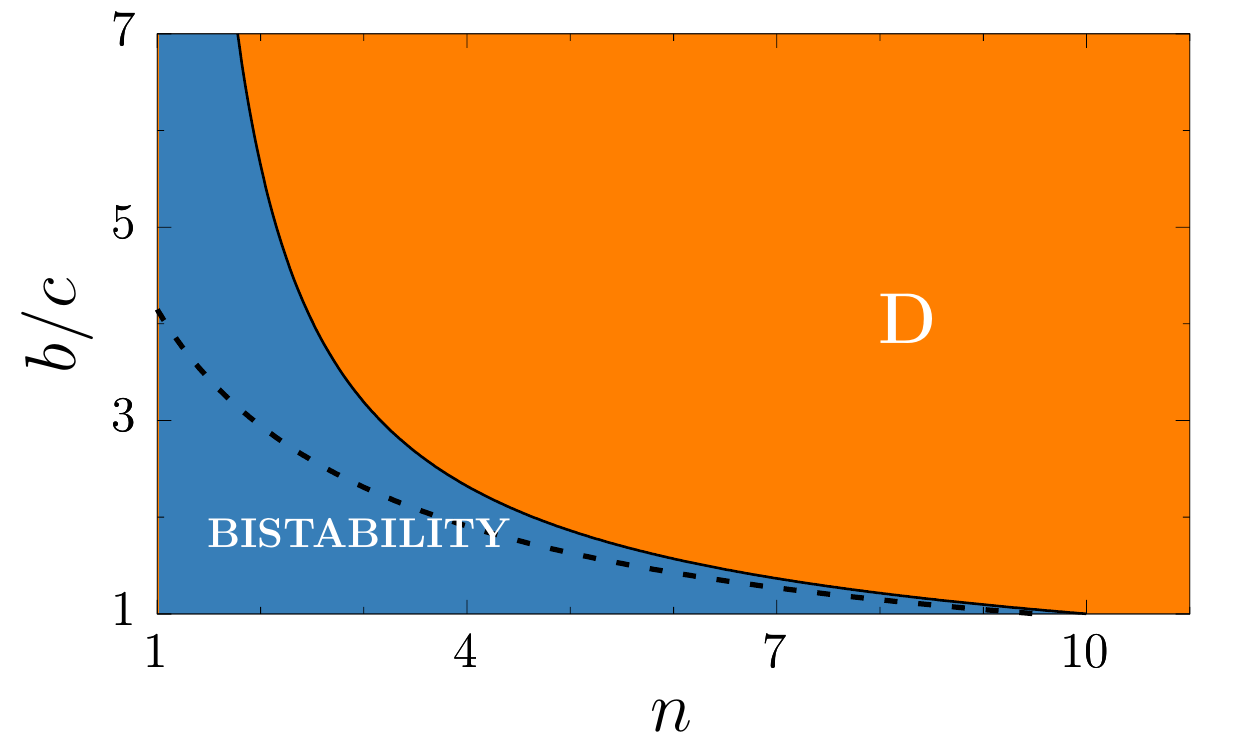}
\caption{Phase diagram for $N_{\max}=100$ and uniform size ($n$) compartments. Only the cooperator-free (D) and the bistable phases are present. Since Eq.~(\ref{eq.boverc}) is valid for integer $n$ only, the solid line separating both phases is just a guide to the eyes. For comparison, the dashed line indicates the lost of stability in the non-uniform case of Fig.~\ref{fig.phasediagram}.}
\label{fig.phasediagram-n}
\end{figure}

When all compartments have the same size $n$, the number of cooperators and defectors respectively become
\begin{align*}
N_{\scriptscriptstyle\rm C}(x) &= x^n N_{\max},\\
N_{\scriptscriptstyle\rm D}(x) &= n^{b/c}  \sum_{m=0} ^{n-1} \!\binom{n}{m} \frac{(1-x)^{n-m} x^m}{(n-m)^{b/c-1}} ,
\end{align*}
and are used to iterate Eq.~(\ref{eq.map}) and to evaluate the stability conditions.
The $x^*=0$ solution is always stable. 
As a consequence, there is neither a pure C phase nor a coexistence one.
For $x^*=1$, the stability condition gives the border between the D and the bistable phase,
\begin{equation}
\frac{b}{c} = \frac{\ln N_{\max}}{\ln n} - 1 .
\label{eq.boverc}
\end{equation}
Since $n$ is a discrete variable, the continuous boundary 
%is a linear-piecewise curve, 
shown in the phase diagram Fig.~\ref{fig.phasediagram-n} is only a guide to the eyes.
For $n=1$, cooperators dominate for all values of $b/c$: isolated cooperators multiply until their number is $N_{\max}$ while defectors remain alone.
Since in this particular case there are no interactions between cooperators and defectors, the value of $b$ is irrelevant.
Similar to Eq.~(\ref{eq.lmax}) for $\lambda_{\max}$, the $x^*=1$ solution exists while  $n\leq\sqrt{N_{\max}}$ when $b/c\to 1$.
For all values of $n$ in this interval there is a threshold value of $b/c$, beyond which cooperators do not survive.
For all these values of $n$, this maximum value is larger than in the non-uniform case (indicated by a dashed line in Fig.~\ref{fig.phasediagram-n}).

Despite of that, size diversity remains very important to cooperation, allowing both the presence of a purely cooperating phase (C) and the coexistence of cooperators and defectors, phases that are absent in the uniform-size case.
%As a tradeoff, it reduces the range of $b/c$ where cooperation may be found.

%%%%%%%%%%%%%%%%%%%%%%%%%%%%%%%%%%%%%%%%%%%%%%
\section{Conclusions}

We introduced a minimal model of a two-level selection mechanism that may sustain cooperative behavior in a population of cooperators and defectors that interact under well-mixed conditions.
More concretely, a large number of finite compartments is created and populated with agents randomly chosen from an infinite pool in which the relative frequency of each strategy is known.
Internally to each compartment, the dynamics is mean-field: agents are well mixed and cooperators only survive in the absence of defectors. 
Reproduction depends on the payoffs obtained through the interactions between agents and the final size of each compartment is related to the amount of cooperators by Eq.~(\ref{eq.replicator}).
Once the asymptotic state is attained in each compartment, they are all merged (the infinite pool) and the density of cooperators is updated.
Besides the within-group competition, the differential growth of diversely composed groups is akin to a reproductive process~\cite{MeCrFr10}.
Thus, different groups compete because some of them grow larger and contribute more to the common pool.
In particular, only defector-free groups contribute to the persistence of cooperators in the population since in mixed groups, defectors dominate.
Our model studies how selection works in fragmented systems and may be useful to understand the role of isolated groups of cooperating individuals as a step in the evolution of multicellular organisms. 
Despite not being the unit of selection, the competition between those groups may correspond to a higher-level, scaffolding mechanism~\cite{DoLaMoRa20,NiBlBoRa23}, setting the conditions for a major evolutionary transition, from single-celled collectives to multicellularity.
In particular, the fragmentation step, with the associated reduced number of agents in each compartment, is akin to the genetic bottleneck that living individuals go through during reproduction, where the small, more homogeneous group of cells chosen to initiate a new individual may help to prevent the inclusion of defective and cheater cells.

%In particular, defector-free compartments increase the amount of cooperators in the infinite pool and, in the next step,  such compartments have a larger probability to be formed.

Cooperation is possible due to  two stochastic processes forming cooperators-only clusters, compartments with different sizes and fluctuations in their composition.
First, the possibility of having small compartments increases the probability of not having defectors in its initial composition (see Ref.~\cite{Eigentler22} for similar experiments with randomness in the initial state).
And second, the diversity in the initial composition of each group, with fluctuating numbers of Cs and Ds, is a fundamental ingredient upon which selection may act, and an important mechanism responsible for cooperators to thrive. 
While there is no dynamics in the infinite pool, during the structured stage, whose timescale is much larger, selection is operational in the compartments.
Inside each of these demes, the agents remain fully mixed and directly compete for the resources provided by the remaining cooperators.
In other words, finiteness of population is implemented by limiting the growth once cooperators are depleted through the replicator equation.
Eventually, defectors fully dominate the compartments in which they were initially present and growth is halted.
%All compartments are then merged, agents mixed and the cycle starts once again.
The disadvantage met by cooperators in the presence of defectors is compensated by the eventual larger cluster of cooperators only.
This mechanism is related to the Simpson paradox and was already observed in some previous works~\cite{HaHoDo06}.

Nowak and May~\cite{NoMa92} were the first to show that network reciprocity, based on persistent, spatially correlated groups of agents, may sustain cooperation in the PDG.
Our results, on the other hand, indicate that even transient groups, under some circumstances, allow a stable population of cooperating agents.
This is remarkable not only because all groups were merged after the growth phase but also because of the well-mixed nature of the within-group dynamics.
Cooperators may dominate the system for all (C phase) or some (bistable phase) of the initial conditions, or coexist with defectors (coexistence phase).
In the latter, even if the fraction of cooperators is a decreasing function of $b/c$, they never disappear if $\lambda$ is small enough, as a consequence of the existence of compartments fully occupied by cooperators in the initial state.
%Secondo me il fatto paradossale ch diminuire il costo della collaborazione finisca con il favorire i defectors vale la pena di essere sottolineato.
It is remarkable that when the cost of cooperation decreases, it is the population of defectors that increases.
When the average compartment size is not too big, $\lambda\leq\ln N_{\max}$, increasing the benefit $b$ helps the defectors and they start to coexist with cooperators ($0<x<1$). Above $\lambda=\ln N_{\max}$, cooperators only survive when the benefit is not too large. 
Whatever the case, increasing the benefit of cooperation (or decreasing its cost) is more advantageous to defectors and jeopardizes the pure cooperating phase (C).
On the other hand, when the average compartment size increases, cooperators may disappear since the coexistence phase is replaced by the defection one (D). 

Birth and death processes are only implicitly taken into account due to the competitive nature of the replicator equation. 
We assume that the timescale to attain the stationary population is much smaller than the time necessary for spontaneous death to become important.
Thus, the merging time of all pools is large enough for all growth processes to attain an stationary state but smaller than the timescales related to microbial death.
The merging step of the dynamics is performed after all compartments have attained their maximum number of individuals. 
However, different timescales may become relevant~\cite{Ryo19} if the compartments coalesce before the stationary state~\cite{CrMeFr12,MeCrFr15}, halting the growth of both cooperators and defectors at a given maximum number (dependent or not on the initial composition or size of the compartment).
Compartments with a final mixture of strategies are thus possible and whether this would increase the fraction of cooperators depends on two competing mechanisms and remains to be checked: although in mixed compartments cooperators may no longer be depleted, defector-free compartments may grow less as well.

In addition, there are further extensions of this work that would be interesting to consider.
Instead of merging all compartments at once, coalescence may happen by multiple intermediate steps.
For example, spatial growth may be a local process~\cite{AnVeRi19}, first merging the neighboring pools~\cite{XuDy17}.
Analogously, compartments may be hierarchically organized. 
Merging, in this case, would be analogous to an increase in temperature for a system with a complex energy landscape.
Other games besides the PDG may also be considered, as the Snowdrift~\cite{SzFa07,Smith82}, by changing the relation between the values of the payoff table.
%Another possibility is to have $N_{\max}$, the final compartment population bound, depending on its initial size. 
%Another possibility is to consider a different fragmentation of the environment during the spatial segregation regime.
%For example, the sizes of the compartments may be sampled from a power-law, instead of a Poisson, distribution 
%The exponent $\tau$ of the power-law controls the frequency and sizes of the initial compartments.
%Being long-tailed, this distribution may have interesting effects, in particular, on the correlation of the size diversity~\cite{MaAzArCo21} and the several phases with different levels of cooperation.
Finally, the well-mixed condition inside each compartment may be replaced by a  spatially structured population, what is known to favor cooperation.

\begin{acknowledgments}
LP acknowledges the hospitality in Porto Alegre where this work was started and JJA is grateful for the hospitality in Rome where it was concluded.
Work partially supported by the Brazilian agencies FAPERGS, FAPERJ, Conselho Nacional de Desenvolvimento Cientí\-fi\-co e Tecnológico (CNPq), and CAPES. 
\end{acknowledgments}

\bibliographystyle{apsrev}   
%\bibliography{bio} 

\begin{thebibliography}{52}
\expandafter\ifx\csname natexlab\endcsname\relax\def\natexlab#1{#1}\fi
\expandafter\ifx\csname bibnamefont\endcsname\relax
  \def\bibnamefont#1{#1}\fi
\expandafter\ifx\csname bibfnamefont\endcsname\relax
  \def\bibfnamefont#1{#1}\fi
\expandafter\ifx\csname citenamefont\endcsname\relax
  \def\citenamefont#1{#1}\fi
\expandafter\ifx\csname url\endcsname\relax
  \def\url#1{\texttt{#1}}\fi
\expandafter\ifx\csname urlprefix\endcsname\relax\def\urlprefix{URL }\fi
\providecommand{\bibinfo}[2]{#2}
\providecommand{\eprint}[2][]{\url{#2}}

\bibitem[{\citenamefont{Axelrod}(1984)}]{Axelrod84}
\bibinfo{author}{\bibfnamefont{R.}~\bibnamefont{Axelrod}},
  \emph{\bibinfo{title}{The Evolution of Cooperation}}
  (\bibinfo{publisher}{BasicBooks}, \bibinfo{address}{New York},
  \bibinfo{year}{1984}).

\bibitem[{\citenamefont{Nowak}(2006)}]{Nowak06b}
\bibinfo{author}{\bibfnamefont{M.~A.} \bibnamefont{Nowak}},
  \bibinfo{journal}{Science} \textbf{\bibinfo{volume}{314}},
  \bibinfo{pages}{1560} (\bibinfo{year}{2006}).

\bibitem[{\citenamefont{Szabó and Fáth}(2007)}]{SzFa07}
\bibinfo{author}{\bibfnamefont{G.}~\bibnamefont{Szabó}} \bibnamefont{and}
  \bibinfo{author}{\bibfnamefont{G.}~\bibnamefont{Fáth}},
  \bibinfo{journal}{Phys. Rep.} \textbf{\bibinfo{volume}{446}},
  \bibinfo{pages}{97} (\bibinfo{year}{2007}).

\bibitem[{\citenamefont{Frey}(2010)}]{Frey10}
\bibinfo{author}{\bibfnamefont{E.}~\bibnamefont{Frey}},
  \bibinfo{journal}{Physica A} \textbf{\bibinfo{volume}{389}},
  \bibinfo{pages}{4265} (\bibinfo{year}{2010}).

\bibitem[{\citenamefont{Perc et~al.}(2013)\citenamefont{Perc, Gómez-Gardeñes,
  Szolnoki, Floría, and Moreno}}]{Percetal2013}
\bibinfo{author}{\bibfnamefont{M.}~\bibnamefont{Perc}},
  \bibinfo{author}{\bibfnamefont{J.}~\bibnamefont{Gómez-Gardeñes}},
  \bibinfo{author}{\bibfnamefont{A.}~\bibnamefont{Szolnoki}},
  \bibinfo{author}{\bibfnamefont{L.~M.} \bibnamefont{Floría}},
  \bibnamefont{and} \bibinfo{author}{\bibfnamefont{Y.}~\bibnamefont{Moreno}},
  \bibinfo{journal}{J. R. Soc. Interface} \textbf{\bibinfo{volume}{10}},
  \bibinfo{pages}{20120997} (\bibinfo{year}{2013}).

\bibitem[{\citenamefont{Adami et~al.}(2016)\citenamefont{Adami, Schossau, and
  Hintze}}]{AdScHi16}
\bibinfo{author}{\bibfnamefont{C.}~\bibnamefont{Adami}},
  \bibinfo{author}{\bibfnamefont{J.}~\bibnamefont{Schossau}}, \bibnamefont{and}
  \bibinfo{author}{\bibfnamefont{A.}~\bibnamefont{Hintze}},
  \bibinfo{journal}{Phys. Life Rev.} \textbf{\bibinfo{volume}{19}},
  \bibinfo{pages}{1} (\bibinfo{year}{2016}).

\bibitem[{\citenamefont{Nowak and May}(1992)}]{NoMa92}
\bibinfo{author}{\bibfnamefont{M.~A.} \bibnamefont{Nowak}} \bibnamefont{and}
  \bibinfo{author}{\bibfnamefont{R.~M.} \bibnamefont{May}},
  \bibinfo{journal}{Nature} \textbf{\bibinfo{volume}{359}},
  \bibinfo{pages}{826} (\bibinfo{year}{1992}).

\bibitem[{\citenamefont{Maynard~Smith}(1982)}]{Smith82}
\bibinfo{author}{\bibfnamefont{J.}~\bibnamefont{Maynard~Smith}},
  \emph{\bibinfo{title}{Evolution and the Theory of Games}}
  (\bibinfo{publisher}{Cambridge University Press},
  \bibinfo{address}{Cambridge, UK}, \bibinfo{year}{1982}).

\bibitem[{\citenamefont{Hofbauer and Sigmund}(1998)}]{HoSi98}
\bibinfo{author}{\bibfnamefont{J.}~\bibnamefont{Hofbauer}} \bibnamefont{and}
  \bibinfo{author}{\bibfnamefont{K.}~\bibnamefont{Sigmund}},
  \emph{\bibinfo{title}{Evolutionary Games and Population Dynamics}}
  (\bibinfo{publisher}{Cambridge University Press}, \bibinfo{year}{1998}).

\bibitem[{\citenamefont{Hauert et~al.}(2006)\citenamefont{Hauert, Holmes, and
  Doebeli}}]{HaHoDo06}
\bibinfo{author}{\bibfnamefont{C.}~\bibnamefont{Hauert}},
  \bibinfo{author}{\bibfnamefont{M.}~\bibnamefont{Holmes}}, \bibnamefont{and}
  \bibinfo{author}{\bibfnamefont{M.}~\bibnamefont{Doebeli}},
  \bibinfo{journal}{Proc. R. Soc. B} \textbf{\bibinfo{volume}{273}},
  \bibinfo{pages}{2565} (\bibinfo{year}{2006}).

\bibitem[{\citenamefont{Tarnita}(2017)}]{Tarnita17}
\bibinfo{author}{\bibfnamefont{C.~E.} \bibnamefont{Tarnita}},
  \bibinfo{journal}{J. Exp. Biol.} \textbf{\bibinfo{volume}{220}},
  \bibinfo{pages}{18} (\bibinfo{year}{2017}).

\bibitem[{\citenamefont{Matsumura et~al.}(2016)\citenamefont{Matsumura, Kun,
  Ryckelynck, Coldren, Szilágyi, Jossinet, Rick, Nghe, Szathmáry, and
  Griffiths}}]{Matsumura16}
\bibinfo{author}{\bibfnamefont{S.}~\bibnamefont{Matsumura}},
  \bibinfo{author}{\bibfnamefont{A.}~\bibnamefont{Kun}},
  \bibinfo{author}{\bibfnamefont{M.}~\bibnamefont{Ryckelynck}},
  \bibinfo{author}{\bibfnamefont{F.}~\bibnamefont{Coldren}},
  \bibinfo{author}{\bibfnamefont{A.}~\bibnamefont{Szilágyi}},
  \bibinfo{author}{\bibfnamefont{F.}~\bibnamefont{Jossinet}},
  \bibinfo{author}{\bibfnamefont{C.}~\bibnamefont{Rick}},
  \bibinfo{author}{\bibfnamefont{P.}~\bibnamefont{Nghe}},
  \bibinfo{author}{\bibfnamefont{E.}~\bibnamefont{Szathmáry}},
  \bibnamefont{and} \bibinfo{author}{\bibfnamefont{A.~D.}
  \bibnamefont{Griffiths}}, \bibinfo{journal}{Science}
  \textbf{\bibinfo{volume}{354}}, \bibinfo{pages}{1293} (\bibinfo{year}{2016}).

\bibitem[{\citenamefont{Blokhuis et~al.}(2018)\citenamefont{Blokhuis, Lacoste,
  Nghe, and Peliti}}]{BlLaNgPe18}
\bibinfo{author}{\bibfnamefont{A.}~\bibnamefont{Blokhuis}},
  \bibinfo{author}{\bibfnamefont{D.}~\bibnamefont{Lacoste}},
  \bibinfo{author}{\bibfnamefont{P.}~\bibnamefont{Nghe}}, \bibnamefont{and}
  \bibinfo{author}{\bibfnamefont{L.}~\bibnamefont{Peliti}},
  \bibinfo{journal}{Phys. Rev. Lett.} \textbf{\bibinfo{volume}{120}},
  \bibinfo{pages}{158101} (\bibinfo{year}{2018}).

\bibitem[{\citenamefont{Blokhuis et~al.}(2020)\citenamefont{Blokhuis, Nghe,
  Peliti, and Lacoste}}]{BlNgPeLa20}
\bibinfo{author}{\bibfnamefont{A.}~\bibnamefont{Blokhuis}},
  \bibinfo{author}{\bibfnamefont{P.}~\bibnamefont{Nghe}},
  \bibinfo{author}{\bibfnamefont{L.}~\bibnamefont{Peliti}}, \bibnamefont{and}
  \bibinfo{author}{\bibfnamefont{D.}~\bibnamefont{Lacoste}},
  \bibinfo{journal}{J. Theor. Biol.} \textbf{\bibinfo{volume}{487}},
  \bibinfo{pages}{110110} (\bibinfo{year}{2020}).

\bibitem[{\citenamefont{Cooney et~al.}(2023)\citenamefont{Cooney, Levin, Mori,
  and Plotkin}}]{CoLeMoPl23}
\bibinfo{author}{\bibfnamefont{D.~B.} \bibnamefont{Cooney}},
  \bibinfo{author}{\bibfnamefont{S.~A.} \bibnamefont{Levin}},
  \bibinfo{author}{\bibfnamefont{Y.}~\bibnamefont{Mori}}, \bibnamefont{and}
  \bibinfo{author}{\bibfnamefont{J.~B.} \bibnamefont{Plotkin}},
  \bibinfo{journal}{Proc. Natl. Acad. Sci.} \textbf{\bibinfo{volume}{120}},
  \bibinfo{pages}{e2216186120} (\bibinfo{year}{2023}).

\bibitem[{\citenamefont{Gilpin}(1994)}]{Gilpin94}
\bibinfo{author}{\bibfnamefont{M.}~\bibnamefont{Gilpin}},
  \bibinfo{journal}{Proc. Natl. Acad. Sci.} \textbf{\bibinfo{volume}{91}},
  \bibinfo{pages}{3252} (\bibinfo{year}{1994}).

\bibitem[{\citenamefont{Rillig et~al.}(2015)\citenamefont{Rillig, Antonovics,
  Caruso, Lehmann, Powell, Veresoglou, and Verbruggen}}]{Rillig15}
\bibinfo{author}{\bibfnamefont{M.~C.} \bibnamefont{Rillig}},
  \bibinfo{author}{\bibfnamefont{J.}~\bibnamefont{Antonovics}},
  \bibinfo{author}{\bibfnamefont{T.}~\bibnamefont{Caruso}},
  \bibinfo{author}{\bibfnamefont{A.}~\bibnamefont{Lehmann}},
  \bibinfo{author}{\bibfnamefont{J.~R.} \bibnamefont{Powell}},
  \bibinfo{author}{\bibfnamefont{S.~D.} \bibnamefont{Veresoglou}},
  \bibnamefont{and}
  \bibinfo{author}{\bibfnamefont{E.}~\bibnamefont{Verbruggen}},
  \bibinfo{journal}{Trends Ecol. Evol.} \textbf{\bibinfo{volume}{30}},
  \bibinfo{pages}{470} (\bibinfo{year}{2015}).

\bibitem[{\citenamefont{Tikhonov}(2016)}]{Tikhonov16}
\bibinfo{author}{\bibfnamefont{M.}~\bibnamefont{Tikhonov}},
  \bibinfo{journal}{eLife} \textbf{\bibinfo{volume}{5}},
  \bibinfo{pages}{e15747} (\bibinfo{year}{2016}).

\bibitem[{\citenamefont{Antonovics et~al.}(2019)\citenamefont{Antonovics,
  Veresoglou, and Rillig}}]{AnVeRi19}
\bibinfo{author}{\bibfnamefont{J.}~\bibnamefont{Antonovics}},
  \bibinfo{author}{\bibfnamefont{S.~D.} \bibnamefont{Veresoglou}},
  \bibnamefont{and} \bibinfo{author}{\bibfnamefont{M.~C.} \bibnamefont{Rillig}}
  (\bibinfo{year}{2019}), \bibinfo{note}{arXiv:1905.03669}.

\bibitem[{\citenamefont{Dutton et~al.}(2021)\citenamefont{Dutton, Subalusky,
  Sanchez, Estrela, Lu, Hamilton, Njoroge, Rosi, and Post}}]{Dutton21}
\bibinfo{author}{\bibfnamefont{C.~L.} \bibnamefont{Dutton}},
  \bibinfo{author}{\bibfnamefont{A.~L.} \bibnamefont{Subalusky}},
  \bibinfo{author}{\bibfnamefont{A.}~\bibnamefont{Sanchez}},
  \bibinfo{author}{\bibfnamefont{S.}~\bibnamefont{Estrela}},
  \bibinfo{author}{\bibfnamefont{N.}~\bibnamefont{Lu}},
  \bibinfo{author}{\bibfnamefont{S.~K.} \bibnamefont{Hamilton}},
  \bibinfo{author}{\bibfnamefont{L.}~\bibnamefont{Njoroge}},
  \bibinfo{author}{\bibfnamefont{E.~J.} \bibnamefont{Rosi}}, \bibnamefont{and}
  \bibinfo{author}{\bibfnamefont{D.~M.} \bibnamefont{Post}},
  \bibinfo{journal}{Sci. Rep.} \textbf{\bibinfo{volume}{11}},
  \bibinfo{pages}{23117} (\bibinfo{year}{2021}).

\bibitem[{\citenamefont{Lechón-Alonso
  et~al.}(2021)\citenamefont{Lechón-Alonso, Clegg, Cook, Smith, and
  Pawar}}]{LeClCoSmPa21}
\bibinfo{author}{\bibfnamefont{P.}~\bibnamefont{Lechón-Alonso}},
  \bibinfo{author}{\bibfnamefont{T.}~\bibnamefont{Clegg}},
  \bibinfo{author}{\bibfnamefont{J.}~\bibnamefont{Cook}},
  \bibinfo{author}{\bibfnamefont{T.~P.} \bibnamefont{Smith}}, \bibnamefont{and}
  \bibinfo{author}{\bibfnamefont{S.}~\bibnamefont{Pawar}},
  \bibinfo{journal}{PLOS One} \textbf{\bibinfo{volume}{17}},
  \bibinfo{pages}{e1009584} (\bibinfo{year}{2021}).

\bibitem[{\citenamefont{Diaz-Colunga et~al.}(2022)\citenamefont{Diaz-Colunga,
  Lu, Sanchez-Gorostiaga, Chang, Cai, Goldford, Tikhonov, and
  Sánchez}}]{Colunga22}
\bibinfo{author}{\bibfnamefont{J.}~\bibnamefont{Diaz-Colunga}},
  \bibinfo{author}{\bibfnamefont{N.}~\bibnamefont{Lu}},
  \bibinfo{author}{\bibfnamefont{A.}~\bibnamefont{Sanchez-Gorostiaga}},
  \bibinfo{author}{\bibfnamefont{C.-Y.} \bibnamefont{Chang}},
  \bibinfo{author}{\bibfnamefont{H.~S.} \bibnamefont{Cai}},
  \bibinfo{author}{\bibfnamefont{J.~E.} \bibnamefont{Goldford}},
  \bibinfo{author}{\bibfnamefont{M.}~\bibnamefont{Tikhonov}}, \bibnamefont{and}
  \bibinfo{author}{\bibfnamefont{A.}~\bibnamefont{Sánchez}},
  \bibinfo{journal}{Proc. Natl. Acad. Sci.} \textbf{\bibinfo{volume}{119}},
  \bibinfo{pages}{e2111261119} (\bibinfo{year}{2022}).

\bibitem[{\citenamefont{Sousa}(1979)}]{Souza79}
\bibinfo{author}{\bibfnamefont{W.~P.} \bibnamefont{Sousa}},
  \bibinfo{journal}{Ecol. Monog.} \textbf{\bibinfo{volume}{49}},
  \bibinfo{pages}{227} (\bibinfo{year}{1979}).

\bibitem[{\citenamefont{Blaustein and Schwartz}(2001)}]{BlSc01}
\bibinfo{author}{\bibfnamefont{L.}~\bibnamefont{Blaustein}} \bibnamefont{and}
  \bibinfo{author}{\bibfnamefont{S.~S.} \bibnamefont{Schwartz}},
  \bibinfo{journal}{Isr. J. Zool.} \textbf{\bibinfo{volume}{47}},
  \bibinfo{pages}{303} (\bibinfo{year}{2001}).

\bibitem[{\citenamefont{Veach et~al.}(2016)\citenamefont{Veach, Stegen, Brown,
  Dodds, and Jumpponen}}]{Veach16}
\bibinfo{author}{\bibfnamefont{A.~M.} \bibnamefont{Veach}},
  \bibinfo{author}{\bibfnamefont{J.~C.} \bibnamefont{Stegen}},
  \bibinfo{author}{\bibfnamefont{S.~P.} \bibnamefont{Brown}},
  \bibinfo{author}{\bibfnamefont{W.~K.} \bibnamefont{Dodds}}, \bibnamefont{and}
  \bibinfo{author}{\bibfnamefont{A.}~\bibnamefont{Jumpponen}},
  \bibinfo{journal}{Mol. Ecol.} \textbf{\bibinfo{volume}{25}},
  \bibinfo{pages}{4674} (\bibinfo{year}{2016}).

\bibitem[{\citenamefont{Xu and Van~Dyken}(2017)}]{XuDy17}
\bibinfo{author}{\bibfnamefont{S.}~\bibnamefont{Xu}} \bibnamefont{and}
  \bibinfo{author}{\bibfnamefont{J.~D.} \bibnamefont{Van~Dyken}},
  \bibinfo{journal}{Evolution} \textbf{\bibinfo{volume}{72}},
  \bibinfo{pages}{153} (\bibinfo{year}{2017}).

\bibitem[{\citenamefont{Nadell et~al.}(2016)\citenamefont{Nadell, Drescher, and
  Foster}}]{NaDrFo16}
\bibinfo{author}{\bibfnamefont{C.~D.} \bibnamefont{Nadell}},
  \bibinfo{author}{\bibfnamefont{K.}~\bibnamefont{Drescher}}, \bibnamefont{and}
  \bibinfo{author}{\bibfnamefont{K.~R.} \bibnamefont{Foster}},
  \bibinfo{journal}{Nat. Rev.} \textbf{\bibinfo{volume}{14}},
  \bibinfo{pages}{589} (\bibinfo{year}{2016}).

\bibitem[{\citenamefont{Nitschke et~al.}(2023)\citenamefont{Nitschke, Black,
  Bourrat, and Rainey}}]{NiBlBoRa23}
\bibinfo{author}{\bibfnamefont{M.~C.} \bibnamefont{Nitschke}},
  \bibinfo{author}{\bibfnamefont{A.~J.} \bibnamefont{Black}},
  \bibinfo{author}{\bibfnamefont{P.}~\bibnamefont{Bourrat}}, \bibnamefont{and}
  \bibinfo{author}{\bibfnamefont{P.~B.} \bibnamefont{Rainey}},
  \bibinfo{journal}{J. Theor. Biol.} \textbf{\bibinfo{volume}{561}},
  \bibinfo{pages}{111414} (\bibinfo{year}{2023}).

\bibitem[{\citenamefont{Griffin et~al.}(2004)\citenamefont{Griffin, West, and
  Buckling}}]{GrWeBu04}
\bibinfo{author}{\bibfnamefont{A.~S.} \bibnamefont{Griffin}},
  \bibinfo{author}{\bibfnamefont{S.~A.} \bibnamefont{West}}, \bibnamefont{and}
  \bibinfo{author}{\bibfnamefont{A.}~\bibnamefont{Buckling}},
  \bibinfo{journal}{Nature} \textbf{\bibinfo{volume}{430}},
  \bibinfo{pages}{1024} (\bibinfo{year}{2004}).

\bibitem[{\citenamefont{Chuang et~al.}(2009)\citenamefont{Chuang, Rivoire, and
  Leibler}}]{ChRiLe09}
\bibinfo{author}{\bibfnamefont{J.~S.} \bibnamefont{Chuang}},
  \bibinfo{author}{\bibfnamefont{O.}~\bibnamefont{Rivoire}}, \bibnamefont{and}
  \bibinfo{author}{\bibfnamefont{S.}~\bibnamefont{Leibler}},
  \bibinfo{journal}{Science} \textbf{\bibinfo{volume}{323}},
  \bibinfo{pages}{272} (\bibinfo{year}{2009}).

\bibitem[{\citenamefont{Wynne-Edwards}(1963)}]{Wynne63}
\bibinfo{author}{\bibfnamefont{V.~C.} \bibnamefont{Wynne-Edwards}},
  \bibinfo{journal}{Nature} \textbf{\bibinfo{volume}{200}},
  \bibinfo{pages}{623} (\bibinfo{year}{1963}).

\bibitem[{\citenamefont{Wilson}(1975)}]{Wilson75}
\bibinfo{author}{\bibfnamefont{D.~S.} \bibnamefont{Wilson}},
  \bibinfo{journal}{Proc. Natl. Acad. Sci.} \textbf{\bibinfo{volume}{72}},
  \bibinfo{pages}{143} (\bibinfo{year}{1975}).

\bibitem[{\citenamefont{Maynard-Smith}(1976)}]{Smith76}
\bibinfo{author}{\bibfnamefont{J.}~\bibnamefont{Maynard-Smith}},
  \bibinfo{journal}{Quart. Rev. Biol.} \textbf{\bibinfo{volume}{51}},
  \bibinfo{pages}{277} (\bibinfo{year}{1976}).

\bibitem[{\citenamefont{Wilson and Wilson}(2008)}]{WiWi08}
\bibinfo{author}{\bibfnamefont{D.~S.} \bibnamefont{Wilson}} \bibnamefont{and}
  \bibinfo{author}{\bibfnamefont{E.~O.} \bibnamefont{Wilson}},
  \bibinfo{journal}{Am. Sci.} \textbf{\bibinfo{volume}{96}},
  \bibinfo{pages}{380} (\bibinfo{year}{2008}).

\bibitem[{\citenamefont{Vainstein et~al.}(2007)\citenamefont{Vainstein, Silva,
  and Arenzon}}]{VaSiAr07}
\bibinfo{author}{\bibfnamefont{M.~H.} \bibnamefont{Vainstein}},
  \bibinfo{author}{\bibfnamefont{A.~T.~C.} \bibnamefont{Silva}},
  \bibnamefont{and} \bibinfo{author}{\bibfnamefont{J.~J.}
  \bibnamefont{Arenzon}}, \bibinfo{journal}{J. Theor. Biol.}
  \textbf{\bibinfo{volume}{244}}, \bibinfo{pages}{722} (\bibinfo{year}{2007}).

\bibitem[{\citenamefont{Meloni et~al.}(2009)\citenamefont{Meloni, Buscarino,
  Fortuna, Frasca, G\'{o}mez-Garde\~{n}es, Latora, and
  Moreno}}]{MeBuFoFrGoLaMo09}
\bibinfo{author}{\bibfnamefont{S.}~\bibnamefont{Meloni}},
  \bibinfo{author}{\bibfnamefont{A.}~\bibnamefont{Buscarino}},
  \bibinfo{author}{\bibfnamefont{L.}~\bibnamefont{Fortuna}},
  \bibinfo{author}{\bibfnamefont{M.}~\bibnamefont{Frasca}},
  \bibinfo{author}{\bibfnamefont{J.}~\bibnamefont{G\'{o}mez-Garde\~{n}es}},
  \bibinfo{author}{\bibfnamefont{V.}~\bibnamefont{Latora}}, \bibnamefont{and}
  \bibinfo{author}{\bibfnamefont{Y.}~\bibnamefont{Moreno}},
  \bibinfo{journal}{Physical Review E} \textbf{\bibinfo{volume}{79}},
  \bibinfo{pages}{067101} (\bibinfo{year}{2009}).

\bibitem[{\citenamefont{Vainstein et~al.}(2014)\citenamefont{Vainstein, Brito,
  and Arenzon}}]{VaBrAr14}
\bibinfo{author}{\bibfnamefont{M.~H.} \bibnamefont{Vainstein}},
  \bibinfo{author}{\bibfnamefont{C.}~\bibnamefont{Brito}}, \bibnamefont{and}
  \bibinfo{author}{\bibfnamefont{J.~J.} \bibnamefont{Arenzon}},
  \bibinfo{journal}{Phys. Rev. E} \textbf{\bibinfo{volume}{90}},
  \bibinfo{pages}{022132} (\bibinfo{year}{2014}).

\bibitem[{\citenamefont{Javarone}(2016)}]{Javarone16}
\bibinfo{author}{\bibfnamefont{M.~A.} \bibnamefont{Javarone}},
  \bibinfo{journal}{Eur. Phys. J. B} \textbf{\bibinfo{volume}{89}},
  \bibinfo{pages}{42} (\bibinfo{year}{2016}).

\bibitem[{\citenamefont{Cremer et~al.}(2012)\citenamefont{Cremer, Melbinger,
  and Frey}}]{CrMeFr12}
\bibinfo{author}{\bibfnamefont{J.}~\bibnamefont{Cremer}},
  \bibinfo{author}{\bibfnamefont{A.}~\bibnamefont{Melbinger}},
  \bibnamefont{and} \bibinfo{author}{\bibfnamefont{E.}~\bibnamefont{Frey}},
  \bibinfo{journal}{Sci. Rep.} \textbf{\bibinfo{volume}{2}},
  \bibinfo{pages}{281} (\bibinfo{year}{2012}).

\bibitem[{\citenamefont{Steiner}()}]{Steiner21}
\bibinfo{author}{\bibfnamefont{K.~F.} \bibnamefont{Steiner}},
  \bibinfo{note}{bioRxiv/2021.02.21.431661}.

\bibitem[{\citenamefont{Huang et~al.}(2015)\citenamefont{Huang, Hauert, and
  Traulsen}}]{HuHaTr15}
\bibinfo{author}{\bibfnamefont{W.}~\bibnamefont{Huang}},
  \bibinfo{author}{\bibfnamefont{C.}~\bibnamefont{Hauert}}, \bibnamefont{and}
  \bibinfo{author}{\bibfnamefont{A.}~\bibnamefont{Traulsen}},
  \bibinfo{journal}{Proc. Natl. Acad. Sci.} \textbf{\bibinfo{volume}{112}},
  \bibinfo{pages}{9064} (\bibinfo{year}{2015}).

\bibitem[{\citenamefont{Simpson}(1951)}]{Simpson51}
\bibinfo{author}{\bibfnamefont{E.~H.} \bibnamefont{Simpson}},
  \bibinfo{journal}{J. R. Stat. Soc. B} \textbf{\bibinfo{volume}{13}},
  \bibinfo{pages}{238} (\bibinfo{year}{1951}).

\bibitem[{\citenamefont{Blyth}(1972)}]{Blyth72}
\bibinfo{author}{\bibfnamefont{C.~R.} \bibnamefont{Blyth}},
  \bibinfo{journal}{J. Am. Stat. Assoc.} \textbf{\bibinfo{volume}{67}},
  \bibinfo{pages}{364} (\bibinfo{year}{1972}).

\bibitem[{\citenamefont{Hauert et~al.}(2002)\citenamefont{Hauert, {de Monte},
  Hofbauer, and Sigmund}}]{HaMoHoSi02}
\bibinfo{author}{\bibfnamefont{C.}~\bibnamefont{Hauert}},
  \bibinfo{author}{\bibfnamefont{S.}~\bibnamefont{{de Monte}}},
  \bibinfo{author}{\bibfnamefont{J.}~\bibnamefont{Hofbauer}}, \bibnamefont{and}
  \bibinfo{author}{\bibfnamefont{K.}~\bibnamefont{Sigmund}},
  \bibinfo{journal}{Science} \textbf{\bibinfo{volume}{296}},
  \bibinfo{pages}{1129} (\bibinfo{year}{2002}).

\bibitem[{\citenamefont{Geyrhofer and Brenner}(2020)}]{GeBr20}
\bibinfo{author}{\bibfnamefont{L.}~\bibnamefont{Geyrhofer}} \bibnamefont{and}
  \bibinfo{author}{\bibfnamefont{N.}~\bibnamefont{Brenner}},
  \bibinfo{journal}{BMC Ecol.} \textbf{\bibinfo{volume}{20}},
  \bibinfo{pages}{14} (\bibinfo{year}{2020}).

\bibitem[{\citenamefont{Maynard~Smith and Szathmary}(1997)}]{SmSz97}
\bibinfo{author}{\bibfnamefont{J.}~\bibnamefont{Maynard~Smith}}
  \bibnamefont{and}
  \bibinfo{author}{\bibfnamefont{E.}~\bibnamefont{Szathmary}},
  \emph{\bibinfo{title}{The Major Transitions in Evolution}}
  (\bibinfo{publisher}{Oxford University Press}, \bibinfo{year}{1997}).

\bibitem[{\citenamefont{Jacobeen et~al.}(2018)\citenamefont{Jacobeen, Pentz,
  Graba, Brandys, Ratcliff, and Yunker}}]{Jacobeen18}
\bibinfo{author}{\bibfnamefont{S.}~\bibnamefont{Jacobeen}},
  \bibinfo{author}{\bibfnamefont{J.~T.} \bibnamefont{Pentz}},
  \bibinfo{author}{\bibfnamefont{E.~C.} \bibnamefont{Graba}},
  \bibinfo{author}{\bibfnamefont{C.~G.} \bibnamefont{Brandys}},
  \bibinfo{author}{\bibfnamefont{W.~C.} \bibnamefont{Ratcliff}},
  \bibnamefont{and} \bibinfo{author}{\bibfnamefont{P.~J.}
  \bibnamefont{Yunker}}, \bibinfo{journal}{Nat. Phys.}
  \textbf{\bibinfo{volume}{14}}, \bibinfo{pages}{286} (\bibinfo{year}{2018}).

\bibitem[{\citenamefont{Melbinger et~al.}(2010)\citenamefont{Melbinger, Cremer,
  and Frey}}]{MeCrFr10}
\bibinfo{author}{\bibfnamefont{A.}~\bibnamefont{Melbinger}},
  \bibinfo{author}{\bibfnamefont{J.}~\bibnamefont{Cremer}}, \bibnamefont{and}
  \bibinfo{author}{\bibfnamefont{E.}~\bibnamefont{Frey}},
  \bibinfo{journal}{Phys. Rev. Lett.} \textbf{\bibinfo{volume}{105}},
  \bibinfo{pages}{178101} (\bibinfo{year}{2010}).

\bibitem[{\citenamefont{Doulcier et~al.}(2020)\citenamefont{Doulcier, Lambert,
  De~Monte, and Rainey}}]{DoLaMoRa20}
\bibinfo{author}{\bibfnamefont{G.}~\bibnamefont{Doulcier}},
  \bibinfo{author}{\bibfnamefont{A.}~\bibnamefont{Lambert}},
  \bibinfo{author}{\bibfnamefont{S.}~\bibnamefont{De~Monte}}, \bibnamefont{and}
  \bibinfo{author}{\bibfnamefont{P.~B.} \bibnamefont{Rainey}},
  \bibinfo{journal}{eLife} \textbf{\bibinfo{volume}{9}},
  \bibinfo{pages}{e53433} (\bibinfo{year}{2020}).

\bibitem[{\citenamefont{Eigentler et~al.}(2022)\citenamefont{Eigentler,
  Kalamara, Ball, MacPhee, and Davidson}}]{Eigentler22}
\bibinfo{author}{\bibfnamefont{L.}~\bibnamefont{Eigentler}},
  \bibinfo{author}{\bibfnamefont{M.}~\bibnamefont{Kalamara}},
  \bibinfo{author}{\bibfnamefont{G.}~\bibnamefont{Ball}},
  \bibinfo{author}{\bibfnamefont{C.~E.} \bibnamefont{MacPhee}},
  \bibnamefont{and} \bibinfo{author}{\bibfnamefont{F.~A.}
  \bibnamefont{Davidson}}, \bibinfo{journal}{The ISME Journal}
  \textbf{\bibinfo{volume}{16}}, \bibinfo{pages}{1512} (\bibinfo{year}{2022}).

\bibitem[{\citenamefont{Ryo et~al.}(2019)\citenamefont{Ryo, Aguilar-Trigueros,
  Pinek, Muller, and Rillig}}]{Ryo19}
\bibinfo{author}{\bibfnamefont{M.}~\bibnamefont{Ryo}},
  \bibinfo{author}{\bibfnamefont{C.~A.} \bibnamefont{Aguilar-Trigueros}},
  \bibinfo{author}{\bibfnamefont{L.}~\bibnamefont{Pinek}},
  \bibinfo{author}{\bibfnamefont{L.~A.} \bibnamefont{Muller}},
  \bibnamefont{and} \bibinfo{author}{\bibfnamefont{M.~C.}
  \bibnamefont{Rillig}}, \bibinfo{journal}{Trends Ecol. Evol.}
  \textbf{\bibinfo{volume}{34}}, \bibinfo{pages}{723} (\bibinfo{year}{2019}).

\bibitem[{\citenamefont{Melbinger et~al.}(2015)\citenamefont{Melbinger, Cremer,
  and Frey}}]{MeCrFr15}
\bibinfo{author}{\bibfnamefont{A.}~\bibnamefont{Melbinger}},
  \bibinfo{author}{\bibfnamefont{J.}~\bibnamefont{Cremer}}, \bibnamefont{and}
  \bibinfo{author}{\bibfnamefont{E.}~\bibnamefont{Frey}}, \bibinfo{journal}{J.
  R. Soc. Interface} \textbf{\bibinfo{volume}{12}}, \bibinfo{pages}{20150171}
  (\bibinfo{year}{2015}).

\end{thebibliography}

\end{document}